\documentclass[traditabstract]{aa}

\usepackage{amsmath}
\usepackage{txfonts}
\usepackage{graphicx}
\usepackage{amssymb}
\usepackage{epstopdf}
\usepackage{natbib}
\usepackage[para]{threeparttable}

\bibpunct{(}{)}{;}{a}{}{,}
\begin{document}

\authorrunning{Shah et. al.}
\titlerunning{Improving GW parameter estimation for compact binaries
  with LISA}

\title{Using electromagnetic observations to aid gravitational-wave
  parameter estimation of compact binaries observed with LISA II:}
\subtitle{The effect of knowing the sky position}

\author{
S.~Shah \inst{\ref{radboud},\ref{nikhef}} \and
G.~Nelemans \inst{\ref{radboud},\ref{nikhef},\ref{leuven}} \and
M.~van~der~Sluys \inst{\ref{radboud},\ref{nikhef}} 
}

\institute{
Department of Astrophysics/ IMAPP, Radboud University Nijmegen, P.O. Box 9010, 6500 GL Nijmegen, The Netherlands, \email{s.shah@astro.ru.nl} \label{radboud} \and
Nikhef – National Institute for Subatomic Physics,  Science Park 105,  1098 XG Amsterdam, The Netherlands \label{nikhef} \and
Institute for Astronomy, KU Leuven, Celestijnenlaan 200D, 3001 Leuven, Belgium \label{leuven} 
}

\date{Received / Accepted }

\abstract
{In this follow-up paper, we continue our study of the effect of using
knowledge from electromagnetic observations in the gravitational
wave (GW) data analysis of Galactic binaries that are predicted to be
observed by the new \textit{Laser Interferometer Space Antenna} in the
low-frequency range, $10^{-4} \:\mathrm{Hz}<f<1 \:\mathrm{Hz}$. 
In the first paper, we have shown that the strong correlation between
amplitude and inclination can be used for mildly inclined binaries to
improve the uncertainty in amplitude, and that this correlation depends
on the inclination of the system. In this paper we investigate the
overall effect of the other orientation parameters, namely the sky
position and the polarisation angle. We find that after the
inclination, the ecliptic latitude of the source has the strongest
effect in determining the GW parameter uncertainties. We ascertain
that the strong correlation we found previously, only depends on the
inclination of the source and not on the other orientation
parameters. We find that knowing the sky position of the source
from electromagnetic data can reduce the GW parameter
uncertainty up to a factor of $\sim 2$, depending on the inclination
and the ecliptic latitude of the system. Knowing the sky position and
inclination can reduce the uncertainty in amplitude by a factor
larger than $40$. We also find that unphysical errors in the
inclinations, which we found when using the Fisher matrix, can affect
the corresponding uncertainties in the amplitudes, which need to be
corrected.
}

\maketitle

\section{Introduction}
Even though population synthesis studies have predicted  
$\sim 10^{8}$ galactic white-dwarf binaries
\cite[e.g.][]{2001A&A...365..491N}, there are only about $\sim 50$
compact binary sources that have been studied at optical, UV, and
X-ray wavelengths \citep[e.g.][]{2010ApJ...711L.138R}. A space-based
gravitational wave (GW) detector, such as the proposed ESA mission
eLISA, is expected to observe millions of these predicted compact
Galactic binaries with periods shorter than about a few hours
\citep{2009CQGra..26i4030N, 2012arXiv1201.3621A}, amongst other
astrophysical sources, and resolve several thousands of these binaries
\citep{2012ApJ...758..131N}. In a recent paper, we investigated the
effect of using knowledge from electromagnetic (EM) observations in
the gravitational wave data analysis of Galactic binaries by exploring
the correlations that might exist between these parameters. We found
that for binaries with relatively low inclinations ($\iota \lesssim 45^{\circ}$), an EM
constraint on inclination can improve the estimate of the GW amplitude
by a significant factor \citep[Paper I,
hereafter]{2012A&A...544A.153S}. The improvement depends on the
signal-to-noise ratio (S/N) of
the source and on the EM constraint on inclination itself. We also found
that this correlation is a strong function of the inclination of the
system, but that it is independent of other angular parameters
that describe the binary, in particular, sky position, and polarisation
angle. From the initial results we also found that several other
correlations that were rather small as a function of inclination
(e.g. Figure 4 in Paper I)
change drastically when computed as a function of sky position. We
address this problem here and also investigate the general dependence
of parameter uncertainties on the orientations of the
binary. Additionally, we discuss the influence of knowing the sky
position from EM data on the parameter uncertainties of the binary sources.  

The angular resolution of cLISA\footnote{classic LISA refers to the
  detector configuration in consideration prior to eLISA (see Paper
  I).} for monochromatic binaries has been studied previously
\citep{1997CQGra..14.1507P, 1998PhRvD..57.7089C,  2006astro.ph..5034R,
  2011PhRvD..83h3006B}. \cite{1997CQGra..14.1507P} have shown 
the uncertainty in ecliptic latitude and ecliptic longitude as a function 
of both the parameters for a specific case of a cross-polarised source 
radiating at $f = 3\mathrm{mHz}$, which would be observable by
cLISA. For a source with
$S/N = 115$, depending on the values of the ecliptic latitude and
longitude, they found the uncertainties in these angles to range from
1 to 8 milliradians.  However, they assumed that except for the sky position, all 
other parameters are known \textit{a priori}. Their study was also
limited by a very simplified  detector response and was tailored for a system at an
inclination which is favourable in extracting its parameters. Relaxing the
assumption of prior knowledge of other parameters, 
\cite{1998PhRvD..57.7089C} has shown that for a binary for a few sets
of sky positions, the solid angle uncertainty ($\sigma_{\Omega}$)
ranges from 3 to 300 square degrees, depending on the frequency of the
source. In subsequent work by \cite{2006astro.ph..5034R}, sky position
uncertainties were estimated for signals modelled for cLISA. They
included chirp as an additional parameter, and these signals were not
limited to a certain polarisation. Recently,
\cite{2011PhRvD..83h3006B} has shown sky position uncertainties  
for various combinations of data channels from cLISA and
 included detailed detector response for the two distinct cases of
low- and high-frequency sources. For a $10^{-3}$ Hz source with an edge-on
orientation and an S/N of 10, an observation of one year is predicted to
give a solid angle resolution of 
$\sim 0.1$ steradians which is a result very similar to the study
by \cite{2006astro.ph..5034R}. However, none of these previous authors provided 
an overall effect of varying multiple angular parameters, which can
have a significant effect on parameter uncertainties. In this paper we
aim to generalise the uncertainties in all parameters for all
possible orientations of the binary and calculate the improvement in
these uncertainties that is possible thanks to prior knowledge of the sky position
and/or inclination of the system. We show this for eLISA. 

A monochromatic binary expected to be observed by eLISA is completely 
described by seven parameters: dimensionless amplitude ($\mathcal{A}$), 
frequency ($f$), polarisation angle ($\psi$), initial GW phase
($\phi_{0}$), inclination ($\cos \iota$), ecliptic latitude
($\sin\beta$), and ecliptic longitude ($\lambda$). 
The inclination of the binary is defined as
$\cos\iota = \vec{n} \cdot \vec{L}$, where $\vec{n}$ is the line-of-sight vector from
the observer to the source and can be described by the sky position
parameters $\beta$ and $\lambda$. $\vec{L}$ is the orientation of the
orbital plane of the binary (or angular momentum vector), which can be
 specified by $\theta_L$ and $\phi_L$. One of the main
findings of Paper I is that the parameter uncertainties of a
monochromatic binary and the three strongest correlations between
these parameters (the correlation between amplitude and inclination
$c_{\mathcal{A}\: \cos \iota}$, the correlation between phase and
polarisation $c_{\phi\psi}$, and the correlation between phase and
frequency $c_{f \phi}$) are strong functions of the source's
inclination $\iota$. Furthermore, we found that the
parameter uncertainties of such a binary also change as a function of
other orientation parameters. These are parameters that describe the
relative geometry of the binary in relation to the detector: 
$\psi$, $\lambda$, and $\beta$. It is natural to then ask the
questions, how the possible strong correlations between the
parameters due to these four orientation parameters influence these
uncertainties, and to which extent knowledge from the 
EM data can improve the remaining parameter determinations. Of the
remaining three parameters, $\mathcal{A}$ (see Eq. 3 in Paper I) is
intrinsic to the source and does not depend on its relative
orientation to the detector.  Since it only scales the S/N of the
system, varying the amplitude thus will affect parameter uncertainties as
expected, where larger $\mathcal{A}$ will give smaller uncertainties
and \emph{vice versa}. The initial phase $\phi_0$, which does not evolve
for a monochromatic binary in the eLISA band, has no effect on the
signal. On the other hand, the effect of frequency on the
parameter accuracies can be significant; the strength of the
modulations in the signal (which is caused by the motion of the
detector around the Sun and relative to the source) affects low-$f$
sources differently than high-$f$ sources. This has been addressed
in detail in previous studies, 
e.g. \cite{2011PhRvD..83h3006B}. Hence, here we focus on the effects on
the parameter uncertainties of a Galactic binary due to sky position
and the orientation parameters.

In this study, we consider low-frequency binaries. We start by
briefly summarising the analysis methods and then describe our Monte
Carlo (MC) simulations in Section 2. In Section 3, we summarise the
important results and our interpretation. In Section 4 we show that
uncertainties can be severely overestimated for parameters with clear
physical bounds, such as inclination, and thus also for parameters that are tightly
correlated with them. We also briefly discuss the S/N limit for which
the results are valid in Section 4, and present our conclusions in
Section 5. 

\section{Monte Carlo simulations}
Our signal and noise modelling and our data analysis method have been
described in Paper I and we briefly summarise them here. The signals from the
binaries that are expected to be observed by eLISA are modelled as
monochromatic sources characterised by seven parameters, as mentioned
above. Most of these sources radiate at low frequencies (milliHertz),
including some of the \emph{verification} binaries like AM~CVn, and 
at these low $f$, the detector response can be derived in the
long-wavelength approximation. The noise is composed of instrumental
noise described as a Gaussian random process, and a foreground noise
mostly arising from the millions of double-detached Galactic
white-dwarf binaries. After removing the binaries, which are predicted
to be resolved due to their strong signals, the foreground noise is also
described by a Gaussian distribution. Given that a signal has a known
waveform and the noise is Gaussian, we can use the Fisher matrix to
estimate the uncertainties in the parameters that describe the
waveform. The inverse of the Fisher information matrix (FIM), known as
the variance-covariance matrix (VCM), gives the uncertainties and the
correlations between the parameters. Given a number of parameters $n$,
one can calculate the $n\times n$ VCM. For details we refer to Paper I. 

Uncertainties in parameters are proportional to the inverse of the
S/N, but they are also influenced by the correlations between
parameters. Generally, a system with higher S/N has smaller parameter
uncertainties and \textit{vice versa}. However, if there is a strong
correlation between any two parameters, this (in addition to the S/N) will
influence their uncertainties.  As mentioned in the previous section,
we find that the overall effect of the sky position and the orientation
parameters of a binary has not been investigated and this is one of
the goals of this study. To do this, we fixed the values of
$\mathcal{A}$, $f$, and $\phi_0$ to those of AM~CVn (see Paper I) and
varied its $\psi$, $\iota$, $\beta$, and $\lambda$ within their
allowed ranges using a Monte Carlo method.  We carried out a
number of of MC simulations. Each simulation corresponds to varying a
different (sub)set of these four parameters simultaneously. The
motivation is to understand how each of the four parameters above
influence the GW data analysis individually and in combination with
each other. We explain these motivations briefly below, which will
become clearer in section 4 where the choice
of each MC simulation is explained in more detail.

First of all, we set the distance of the AM~CVn system close
enough so that the S/Ns are sufficient to believe the Fisher matrix
results (see section 4). The first set of simulations we performed for an
AM~CVn of a fixed $\mathcal{A}$, $f$, and $\phi_0$, while its position
and orientation parameters $\lambda$, $\beta$, $\psi$, and $\iota$
were all varied simultaneously 4000 times. Thus, from the sample of these
4000 systems with various orientation parameters we first 
ascertained that the dominant effect on the parameter uncertainties is
indeed the inclination of the system (result from Paper I). Next, we
can quantify how the other three angles affect the parameter
uncertainties globally. We performed another set of calculations where
we fixed inclination (to a few choices explained later), frequency, and
the phase (like above), and we varied the amplitude (by changing the
distance, $d$), the sky position, and the polarisation angle. For each
choice of the inclination, MC simulations were performed for 500 systems 
to quantify the effect of varying the S/N (by changing the
distance) and the ecliptic latitude, longitude, and the polarisation angle on 
the parameter uncertainties. Finally, to quantify the
possible improvement in the parameter uncertainties gained by prior
(EM) knowledge of the sky location, we carried out 500 MC simulations 
for an AM~CVn at a fixed inclination and varied the rest of the
orientation parameters and the distance. However, instead of a full $7
\times 7$ VCM, we calculated a $5 \times 5$ VCM, where for each system
the randomly chosen sky locations were fixed to that value. Similarly, we
repeated this exercise to quantify the improvement in the parameter
accuracies gained by EM knowledge of inclination in addition to sky
position, and thus we calculated a $4 \times 4$ VCM for each system. For
clarity, we briefly summarise these four sets of MC simulations below: 
\begin{enumerate}
\item MC1: Fix $\mathcal{A}$ (i.e. $d$, $m_1$, and $m_2$ where, $m_i$
  is the mass), $f$, and $\phi_0$. Randomly pick the orientation of
  the binary, $\vec{L}  (\phi_L,\theta_L )$,  $\lambda$, and $\beta$ over their
  physical ranges. The parameters $\psi$, and $\iota$ are then given by the following
  equations \citep{2003PhRvD..67b2001C}:
\begin{equation}
\cos\iota = \cos(\theta_L) \; \sin(\beta) + \sin(\theta_L) \;
\cos(\beta) + \cos(\phi_L-\lambda)  \:\; \mathrm{and} 
\end{equation}
\begin{equation}
\tan\psi = \frac{\sin\beta \; \cos(\lambda-\phi_L) - \cos(\theta_L)\;
  \sin(\theta_L)}{\sin(\theta_L) \; \sin(\lambda-\phi_L)}.
\end{equation}
\item MC2: For each $\iota = 30^{\circ}, 60^{\circ}, 90^{\circ}$, randomly
  pick $\lambda$, $\beta$, $\psi$, and $d$.
\item MC3: Like MC2, but fix $\lambda$, and $\beta$ assuming
  an \textit{a priori} known sky position [$5\times5$ FIM].
\item MC4: Like MC2, but fix $\lambda$, $\beta$, and $\iota$ assuming
  an \textit{a priori} known sky position and inclination [$4\times4$ FIM].
\end{enumerate}
Each of the four parameters are randomly picked from their uniform
distribution in the following range: $\lambda \in [0, 2\pi]$,  $\sin\beta \in [-1,
1]$,  $\phi_L \in [0, 2\pi]$,  $\theta_L \in [0, \pi]$, and $d \in[0.01,
0.3]$ kpc. The corresponding ranges in $\iota$ and $\psi$ are $[0,
\pi]$. We also define the following inclinations for presenting the
results below (unless specified): face-on:= $ 0^{\circ} \leq \iota
\leq 45^{\circ}$ or $135^{\circ} \leq \iota \leq 180^{\circ}$, mildly
face-on:= $45^{\circ} \leq \iota \leq 60^{\circ}$ or $120^{\circ} \leq
\iota \leq 135^{\circ}$, mildly edge-on:= $60^{\circ} \leq \iota \leq
80^{\circ}$ or $100^{\circ} \leq \iota \leq 120^{\circ}$, and edge-on:=
$80^{\circ} \leq \iota \leq 100^{\circ}$. 

\begin{figure}[!h]
\centering
\includegraphics[width=\columnwidth]{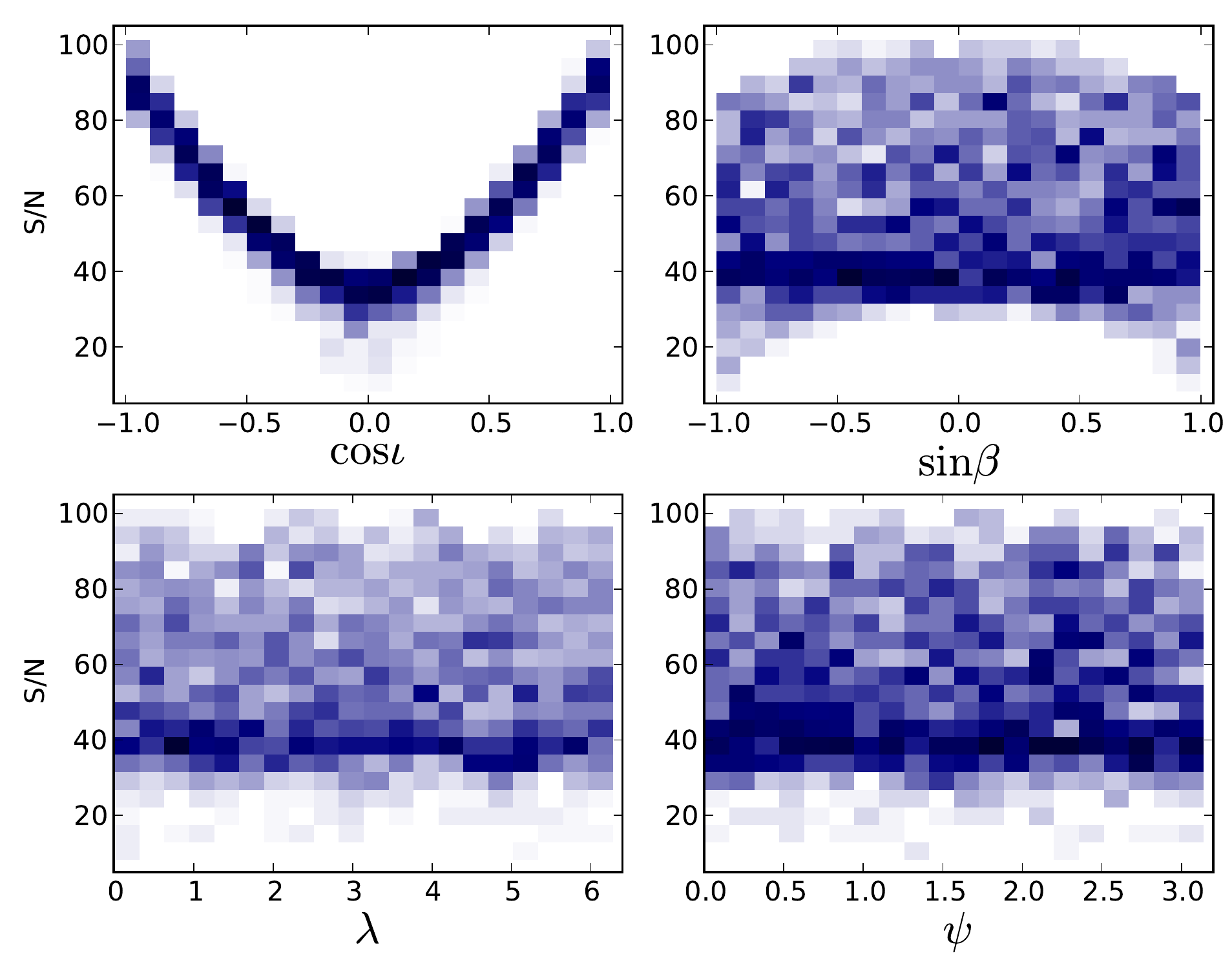}
\caption{\label{fig:SN_params}Two-dimensional distributions of signal-to-noise
  ratio as a function of $\cos\iota$, 
  $\sin\beta$, $\lambda$, and $\psi$. Face-on, mildly face-on, mildly edge-on, and edge-on systems (whose exact bounds are specified in Section 2) can be
  read off from the top-left panel. In the other panels, the
  trend from face-on to edge-on systems follows from the top of the
  distribution to the bottom. }
\end{figure}

\section{Results}
\subsection{Global dependence of the uncertainties: dominance of inclination (MC1)}
We have seen in Paper I that the parameter uncertainties are a strong 
function of the inclination of the system, which is due to the strong
dependence of the signal strength, i.e. S/N, on the inclination and
 to the strong correlations. In this section, we first discuss the
global behaviour of the parameter uncertainties as a result of varying
the sky position, inclination, and the polarisation angle at the same
time. We verify the result from Paper I that inclination has a
dominant effect on the parameter uncertainties. We show 
that after the inclination, the ecliptic latitude,
$\beta$, of the system has the strongest influence on the parameter
uncertainties.

\begin{figure*}
\centering 
\includegraphics[width=18cm]{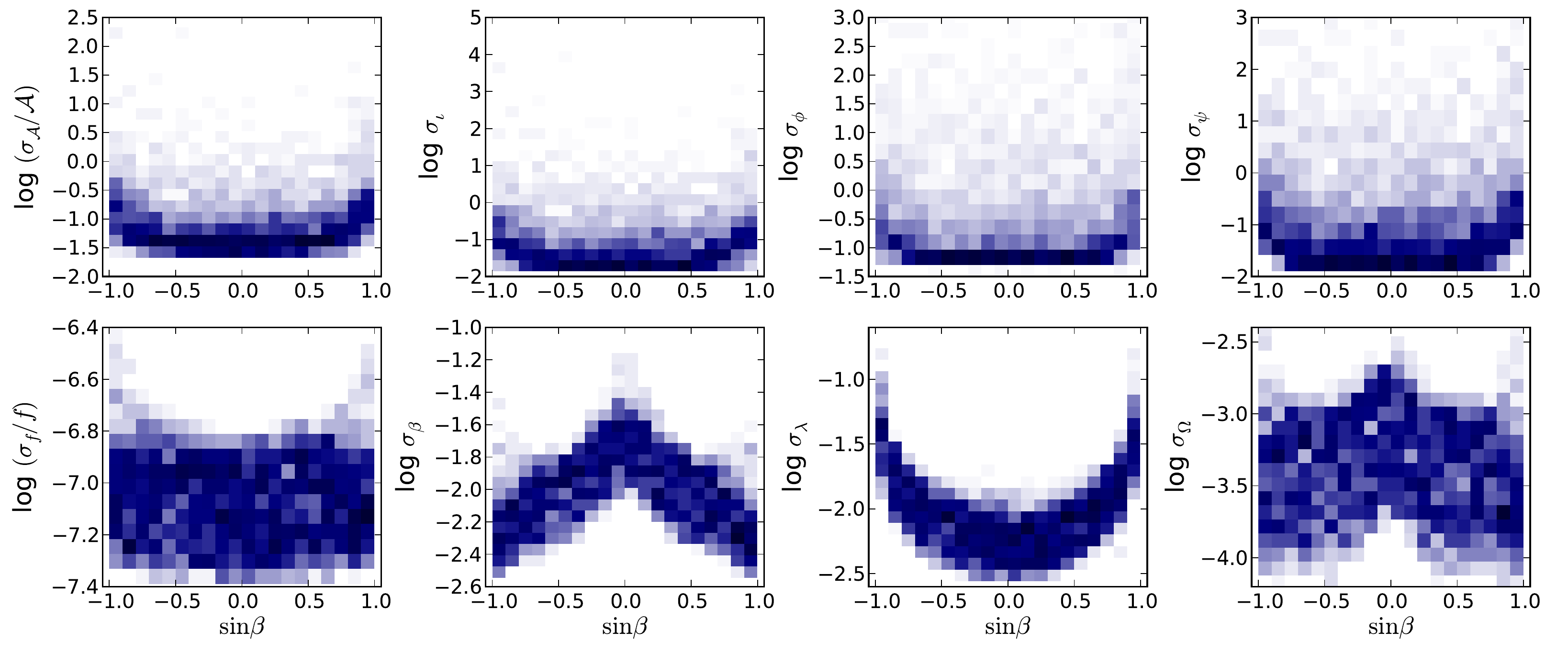}
\caption{Two-dimensional distributions of the uncertainties in the
  seven GW parameters of the galactic binary systems as a function of
  $\sin \beta$ from MC1 as described in the text. Units of the angles are
  in radians. } 
\label{fig:errs_beta}
\end{figure*}

Using the results from MC1, we show in
Figure~\ref{fig:SN_params} the S/N of binary systems as a function of
the four angular parameters over which they are randomly picked. As
expected, there is a strong dependence of S/N of a system on 
inclination, as shown in the top-left panel. Due to this dependence on the
inclination, in the other panels we find that the face-on systems
form a band at the top of these plots, followed by bands of mildly
face-on, mildly edge-on, and edge-on systems, in that order. In the
top-right panel, there is a mild but noticeable increase in S/N, for 
the binaries at the ecliptic plane, $\sin\beta=0$, whereas the
ecliptic longitude and the polarisation angle have no significant effect
on the binaries' S/N, as shown in the bottom panels.
Figure~\ref{fig:errs_beta} is a plot of parameter uncertainties as a
function of $\sin\beta$ taken from MC1. In the top panels, 
uncertainties in $\mathcal{A}$, $\iota$, $\phi$, and $\psi$ are
smallest for edge-on systems (forming a band at the bottom) and 
progressively increase for the face-on systems (the top of
distribution). This is discussed in Paper I and results from the
correlations between these parameters. The lower inclination systems
$\iota \lesssim 45^{\circ}$ or $\iota  \gtrsim 135^{\circ}$ have very
similar signal shapes in those ranges, whereas systems with close to edge-on
orientations are distinguishable by both the shape and structure for
small differences in inclinations. For face-on
systems a small change in inclination is indistinguishable from a
small change in its intrinsic amplitude, whereas for an
edge-on system, a small change in inclination produces a noticeably
different signal. Thus even for their lower S/Ns, the inclinations of
the edge-on systems are better determined than those of the
face-on systems. A similar argument based on the strong correlation
between $\phi$ and $\psi$ for face-on systems can explain the huge
uncertainties of these parameters despite their strong signals (see
Paper I). Since there are no such correlations between $f$, $\lambda$,
and $\beta$ (or the solid angle $\Omega$), their uncertainties should
be a consequence of the S/N, i.e. from the top panels of
Figure~\ref{fig:SN_params}: a higher S/N implies lower values for the
uncertainties and \textit{vice versa}, which indeed holds for
$\sigma_{f}$ and $\sigma_{\lambda}$. From Figures~\ref{fig:SN_params}
and~\ref{fig:errs_beta} it is evident that inclination has the
strongest effect on the S/N and so to the leading order this parameter
affects the parameter uncertainties, which confirms our results in Paper I. 

After inclination, the strongest effect on the S/N of the systems is
produced by the ecliptic latitude.  Thus, we expect the uncertainties in 
the parameters also to be affected by $\beta$ (see top-right panel in
Figure~\ref{fig:SN_params}); sources at the
ecliptic poles should have larger uncertainties than those 
at the ecliptic plane. In Figure ~\ref{fig:errs_beta} we can 
see that for a fixed distance to the source $d$, most 
uncertainties ($\sigma_{\mathcal{A}}$, $\sigma_{\iota}$, $\sigma_{\phi}$,
$\sigma_{\psi}$, $\sigma_{f}$, and $\sigma_{\lambda}$) are indeed smaller 
towards the plane than close to the poles. This is
probably largely due to the S/N as a function of $\sin\beta$.
The unusual behaviour of $\sigma_{\beta}$, which increases at the
plane compared to its value at the poles 
($\sin\beta = \pm 1$), has been pointed out in
\cite{PhysRevD.67.103001} and is caused by the Doppler modulation
$\Phi_{\mathrm{D}}$ (Eq. 9 in Paper I). This behaviour is surprising
since $\Phi_{\mathrm{D}}$ has the largest magnitude on the plane
(since it is proportional to $\cos\beta$), and so we naively expect the sky
resolution to be better on the ecliptic plane. However, the resolution
in $\beta$ is  determined by the \emph{rate of change} of the Doppler
modulation with respect to $\beta$ (i.e. proportional to $\sin\beta$), which is
the highest at the pole and lowest at the plane. This is also evident 
  in the solid angle uncertainty $\sigma_{\Omega}$ in the bottom-right
  panel in Figure~\ref{fig:errs_beta}, where $\sigma_{\Omega} =
  2\pi(\sigma_{\sin\beta}\sigma_{\lambda} -
  \mathcal{C}_{\sin\beta\lambda})$, $\mathcal{C}_{\sin\beta\lambda}$
  being the unnormalised correlation between $\sin\beta$ and $\lambda$
  \citep{1998PhRvD..57.7089C}. 
We also observe that
$\sigma_{\lambda}$ has a distinct U shape, becoming very severe at
the poles. This is merely because $\lambda$ becomes
progressively less well-defined as a source lies closer to the
pole. Note that at the pole $\lambda$ is not defined, which means that
it can take
any value.  
\begin{figure*}
\centering 
\includegraphics[width=18cm]{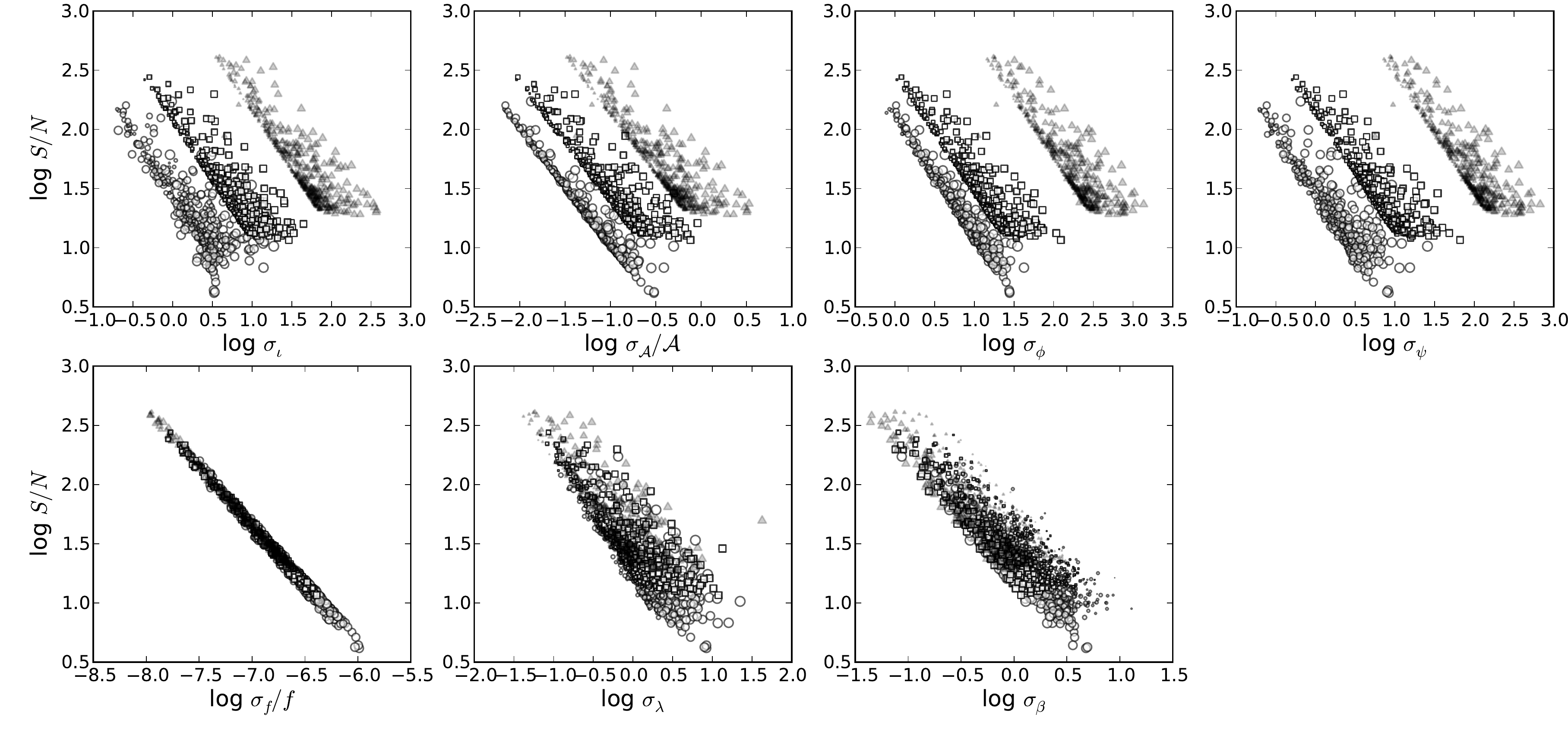}
\caption{ Uncertainties in all seven parameters of the systems as a
  function of S/N from MC2. Open circles, squares, and filled triangles represent
  binary systems at fixed inclinations of $90^{\circ}, 60^{\circ},$ and $30^{\circ}$
  respectively. The size of the symbol represents the magnitude of
  $|\beta|$ for each simulated binary. In the bottom panels
    the different groups of symbols overlap.}  
\label{fig:errs_beta_fix_inc}
\end{figure*}
\begin{figure*}
\centering 
\includegraphics[width=18cm]{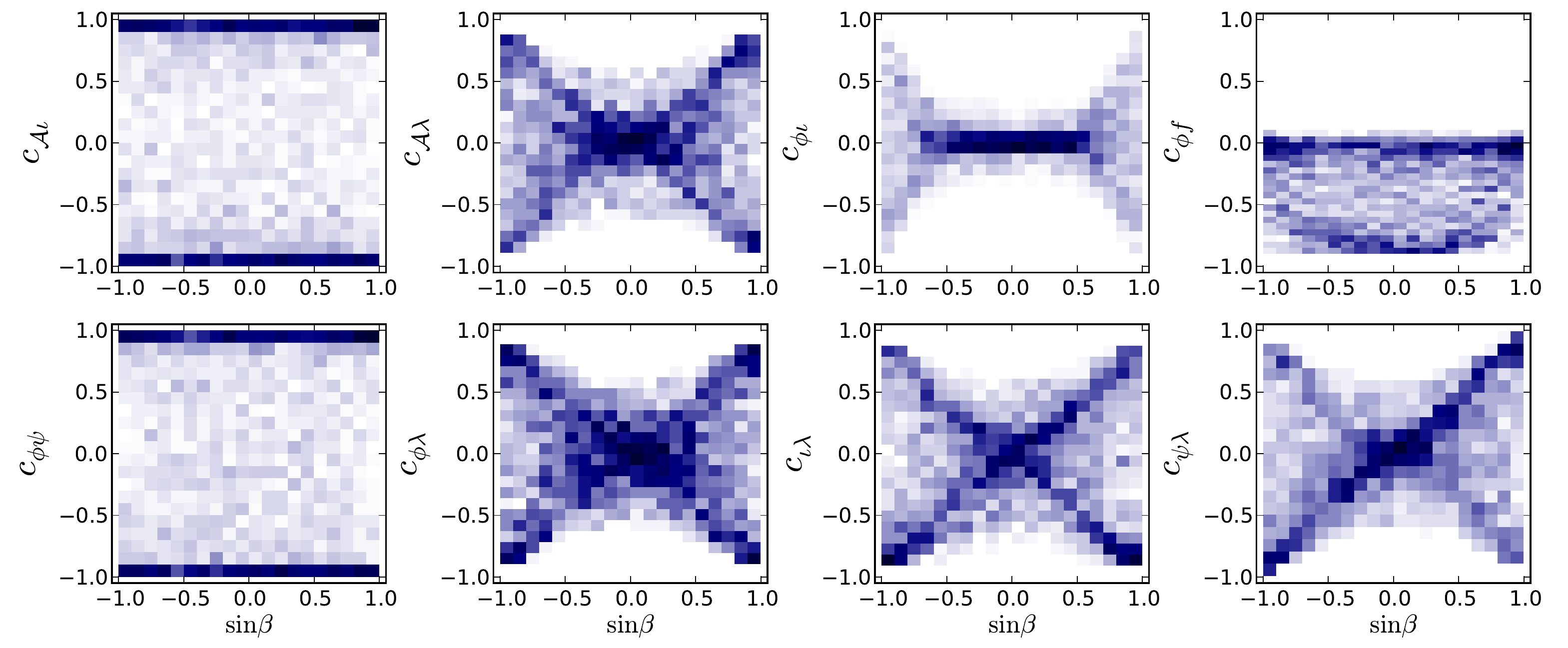}
\caption{\label{fig:corrs_vs_elat} Two-dimensional distributions of
  the normalised correlations as a function of $\sin\beta$ from
  MC2. We show here only the strong (normalised) correlations, meaning $|c_{ij}| > 0.5$. } 
\end{figure*}

\subsection{Dependence of the uncertainties on $\beta$: S/N or
  correlations (MC2)?} 
The general behaviour of the parameter uncertainties (except
$\sigma_{\beta},$ hence $\sigma_{\Omega}$) in
Figure~\ref{fig:errs_beta} shows  
a minimum at the ecliptic plane that could be attributed to the S/N,
since it is the highest for sources in the plane. To investigate this
change in the uncertainties as a
function of $\sin \beta$, we consider the results from MC2 and
the correlations between parameters from MC1. In MC2 we varied S/N (by
varying the distance to the binary, $d$), $\lambda$, $\beta$, and $\psi$ for a few fixed
inclinations. We fixed the $\iota$ for each of the MC2 simulations because we
established its dominant effect through the S/N in the previous
section, therefore the goal now is to determine which contributes most to
the behaviour observed in Figure~\ref{fig:errs_beta}: the S/N (top-right
panel of Figure~\ref{fig:SN_params}), or the correlations as a function
of $\sin \beta$. If the larger uncertainties at the pole compared to
the plane are solely due to the corresponding S/Ns, then we should see from the MC2
simulations that uncertainties anti-correlate tightly
with the S/N of the sources. In
 Figure~\ref{fig:errs_beta_fix_inc} we show 
three distinct curves in S/N vs. parameter uncertainty that roughly show
monotonic functions where all the uncertainties decrease with
increasing S/N. However, all uncertainties except $\sigma_f$ have
a significant horizontal scatter. In the figure, the size of the
symbol represents the magnitude of the ecliptic latitude. Higher
values of $|\beta|$, which represent regions around both ecliptic poles,
concentrate towards the right-hand side of the plots (with the exception of
$\sigma_{\beta}$), and the opposite is true for the lower values of
$|\beta|$, which roughly represent the ecliptic plane. In the top
panels, from left to right, the curves correspond to
$\iota=90^{\circ}, 60^{\circ}$, and $30^{\circ}$ respectively, and these
overlap in the bottom panels. As explained above,  
$\sigma_{\mathcal{A}}$, $\sigma_{\iota}$, $\sigma_{\psi}$, and
$\sigma_{\phi}$ are larger for face-on systems ($\iota=0, 30^{\circ}$) than for
edge-on systems ($\iota=90^{\circ}$), owing to the
correlations between these parameters even though face-on
systems have higher S/Ns. Since these correlations are
very weak for $\lambda$, $\beta$, and $\psi$ for all 
inclinations (see Paper I), their uncertainties overlap
in the bottom panels of Figure~\ref{fig:errs_beta_fix_inc}.
This spread across the horizontal axis must be due to changes in their
correlations as a function of the ecliptic latitude. 
This is indeed shown in Figure~\ref{fig:corrs_vs_elat}, which is
taken from MC1. In the
figure we show a 2d histogram to display the global dependency of the
correlations on the ecliptic latitude. 
We notice that the correlations involving $\lambda$, $\iota$, $\psi$,
$\phi$, and $\mathcal{A}$ ($c_{\mathcal{A}\lambda}$, $c_{\phi\iota}$,
$c_{\phi\lambda}$, $c_{\iota\lambda}$, and $c_{\psi\lambda}$) are
generally weaker at the ecliptic plane. However, at the poles these
correlations can take any value from 0 to $\pm1$ and
  concentrate at $\pm1$. This is the case for
all inclinations. Accordingly, close to the ecliptic poles, there are sets 
of $\lambda$ and $\psi$ for which the correlations can become very
strong, which explains the horizontal 
spread of uncertainties in Figure~\ref{fig:errs_beta}.  Note that in
Figure~\ref{fig:corrs_vs_elat}, we have shown (only) the normalised
correlations $c_{ij}$, whose absolute values can exceed 0.5.
A detailed explanation of these correlations can be found in the
Appendix.

\subsection{ Gain from a priori knowledge of sky position and
  inclination (MC and MC4)}
\begin{figure*}
\centering 
\includegraphics[width=17cm]{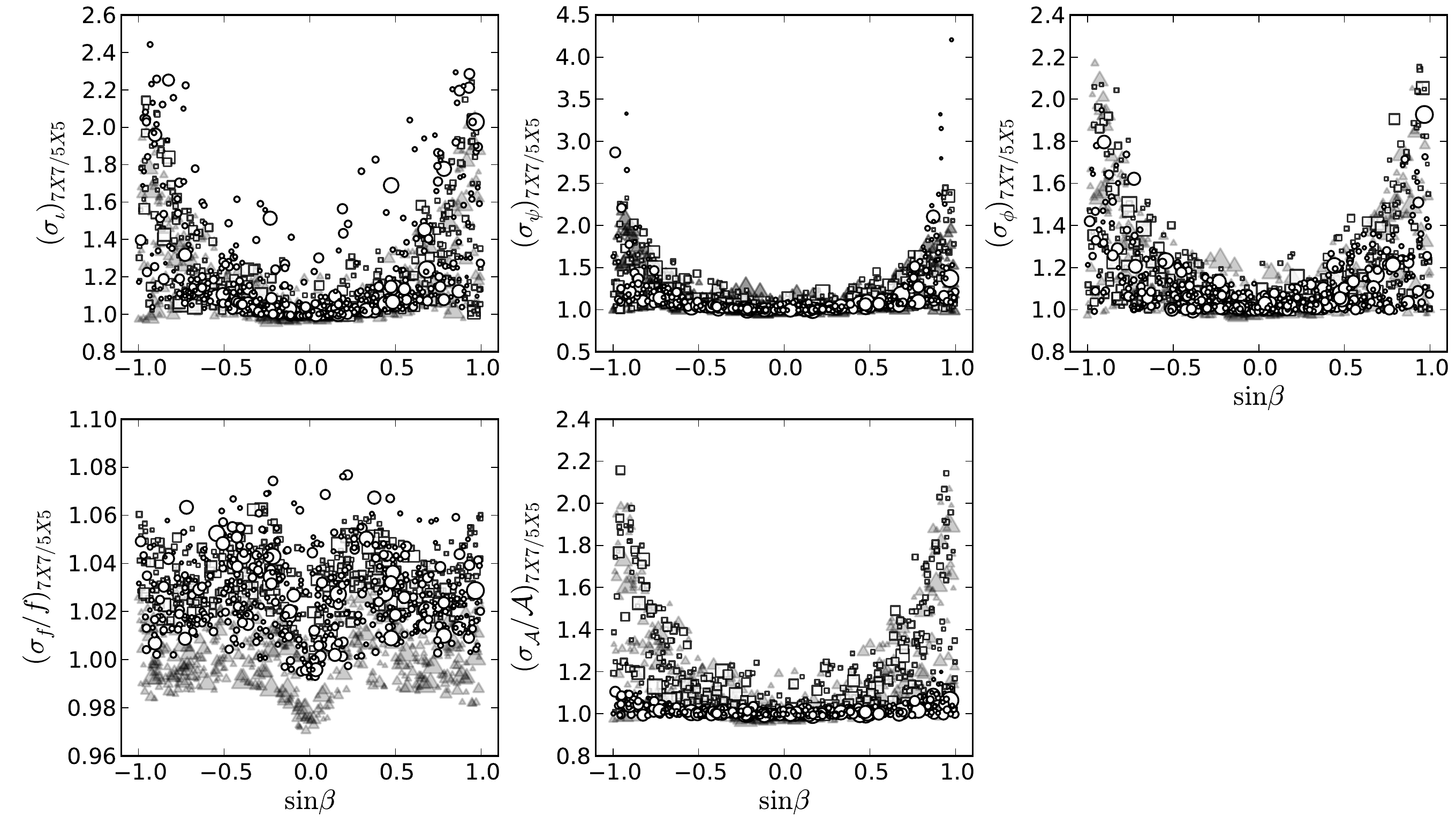}
\caption{Uncertainty ratios of the parameters between analyses using a
  $7\times 7$ VCM
  and a $5\times 5$ VCM, where the sky position is fixed in the latter,
 as a function of $\sin \beta$ (from the 
  MC2 and MC3 simulations). Filled triangles, squares and open circles represent systems
  at fixed inclinations of $30^{\circ}, 60^{\circ}$ and $90^{\circ}$
  respectively. The size of the symbol for each binary represents the S/N
of the system.} 
\label{fig:7X7_5X5}
\end{figure*}

From Figures~\ref{fig:errs_beta} and~\ref{fig:errs_beta_fix_inc} we
observe that uncertainties in sky location are rather large (which has
also been pointed out in previous studies). For the S/Ns in this study
($\sim 8 -100$), the uncertainties in $\sigma_{\Omega}$ range from
$\sim 0.2$ to $6$ square degree, as shown in
Figure~\ref{fig:SN_params}. If EM data can provide a more accurate sky 
position and possibly an inclination for AM~CVn system,  we can use these 
independent (EM) measurements to improve accuracies in other
parameters. To see by which factors the uncertainties improve, we fixed
the sky position of each source and calculated the corresponding
$5\times5$ VCM. The ratios of uncertainties from using a $7\times7$
VCM to using a $5\times5$ VCM, i.e. the improvement when knowing the
sky position (ecliptic latitude and longitude) exactly, are shown in
Figure~\ref{fig:7X7_5X5}, calculated from MC2 and MC3. In the figure,
the size of the symbol represents the S/N of that system. The figure
shows that the improvement factor depends on the inclination of the
source: for edge-on systems (in open circles with $\iota = 90^{\circ}$)
$\sigma_{\psi}$ can improve by a factor of $\sim 3$, but for systems
closer to a face-on configuration in squares and filled triangles the improvement
can only be up to a factor of $\sim 2.2$. One useful result from this
figure is that $\sigma_{\mathcal{A}}$ and  $\sigma_{\iota}$ can
improve by a factor of $\sim 2$ for the systems close to the ecliptic
poles. This gives a motivation for independent EM observations of the
sky location. However, for edge-on systems there is no
improvement in $\sigma_{\mathcal{A}}$ even though $\sigma_{\iota}$
improves\footnote{Depending on the precision of the GW accuracy of the
  inclination, this may aid in finding eclipsing binaries.}. This is
simply due to the lack of correlation between these two parameters for
$\iota = 90^{\circ}$.  
\begin{figure*}
\centering
\includegraphics[width=11cm]{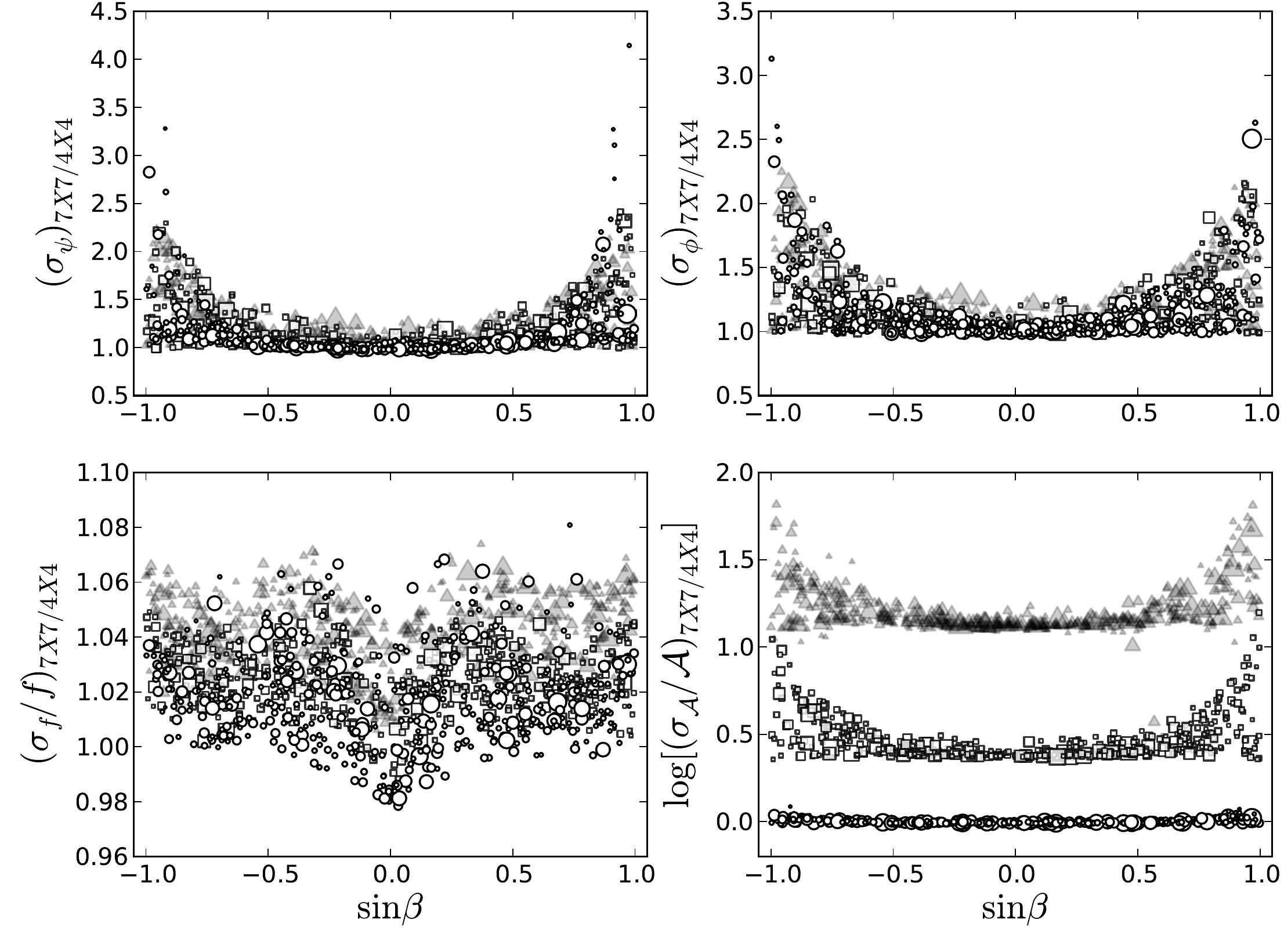}
\caption{\label{fig:7X7_4X4}Same as in Figure~\ref{fig:7X7_5X5} but now we assume that in
  addition to sky position we also know the inclination of the
  binary. The uncertainty ratios of the parameters are between
  $7\times 7$ VCM (MC2) and $4\times 4$ VCM (MC4). The improvement
  factors in amplitude include corrections due to overestimated uncertainties in the
    amplitudes of face-on systems (see Section 4.1) Note that
      the last panel is plotted in log scale in order to highlight the  
      relatively large improvements for the systems with smaller inclinations.}
\end{figure*} 

In Paper I we found that for the HM~Cnc system, the improvement of
$\sigma_{\mathcal{A}}$ by using the EM constraint in $\iota$ was
smaller than for AM~CVn, although the two have similar
inclinations (thus the same (normalised) correlation between 
$\mathcal{A}$ and $\iota$). This was due to the relatively high S/N of
HM~Cnc compared to the latter and thus the individual improvement
factors shown in Figures~\ref{fig:7X7_5X5} and ~\ref{fig:7X7_4X4} will
also depend on the individual S/N of the systems (see Paper I).

If in addition to sky position the inclination can be constrained, as 
discussed previously in Paper I, the improvement factor in amplitude can
be much larger because of the decoupling of the correlation between
amplitude and inclination. We show the ratio of parameter
uncertainties from calculating the $7\times7$ VCM to the
uncertainties from a $4\times4$ VCM (where $\lambda,\: \beta$ and
$\iota$ are fixed) in Figure~\ref{fig:7X7_4X4}. The improvements in
$\sigma_{\psi}$ and $\sigma_{\phi}$ are roughly similar to the case 
in Figure~\ref{fig:7X7_5X5}.  A remarkable difference is seen in the
improvement of $\sigma_{\mathcal{A}}$ for binaries with $\iota =
30^{\circ}$ represented in filled triangles plotted in the bottom-right
panel of Figure~\ref{fig:7X7_4X4}. This is expected because by fixing
$\iota$, the strong correlation between amplitude and 
inclination is decoupled. Hence for the higher S/Ns of close to face-on
systems, the uncertainties in $\mathcal{A}$ is significantly reduce. Note
that here we assumed that the inclination is known exactly.

\section{Discussion}
\subsection{Limiting inclinations to their 
  physical range}
Some of the uncertainties in $\iota$ in Figure~\ref{fig:errs_beta}
from MC1 are noteworthy in that they exceed their physical bounds, for
example $\sigma_{\iota}  = 7 $ radians. The cause for this is that 
the linearised signal approximation that is internally employed in the
Fisher matrix studies is not sensitive to the bounded\footnote{Note that 
    choosing $\cos\iota$ instead of $\iota$ does not alleviate this
    problem since $\cos\iota$ is also a bounded quantity and there is no
    formalism in FIM to handle bounded quantities, whether cyclic
    or not. In fact, for the FIM calculations we evaluate
    uncertainties in $\cos\iota$ and $\sin\beta$, while in quoting our
    results they are converted to uncertainties in $\iota$ and $\beta$ by scaling them as 
  $\sigma_{\iota}=\sigma_{\cos\iota}/\sin\iota$ and
  $\sigma_{\beta}=\sigma_{\sin\beta}/\cos\beta$,  
 since these quantities are more intuitive.} parameters that
describe the model signal, as has been pointed out in
\cite{2008PhRvD..77d2001V}. Thus, when an uncertainty in a (cyclicly) bounded
parameter like the inclination exceeds its physically allowed range, it means the quantity
cannot be determined from GW data analysis alone. For
the parameter $\iota$, we can further ask what the unphysical
uncertainty in $\iota$ implies for the uncertainty in $\mathcal{A}$ for
the mildly inclined systems that has a strong correlation between the
two parameters. Since the Fisher matrix estimates $\sigma_{\iota}$ to unphysical
values while in reality it is constrained within $[0^{\circ}, 90^{\circ}]$, does this mean that
the uncertainty in the amplitude (due to the correlation
with $\iota$) is also overestimated? Or should we not trust the uncertainty
in all those parameters where the $\sigma_{\iota}$ becomes unphysical?
\begin{figure*}
\centering
\includegraphics[width=18cm]{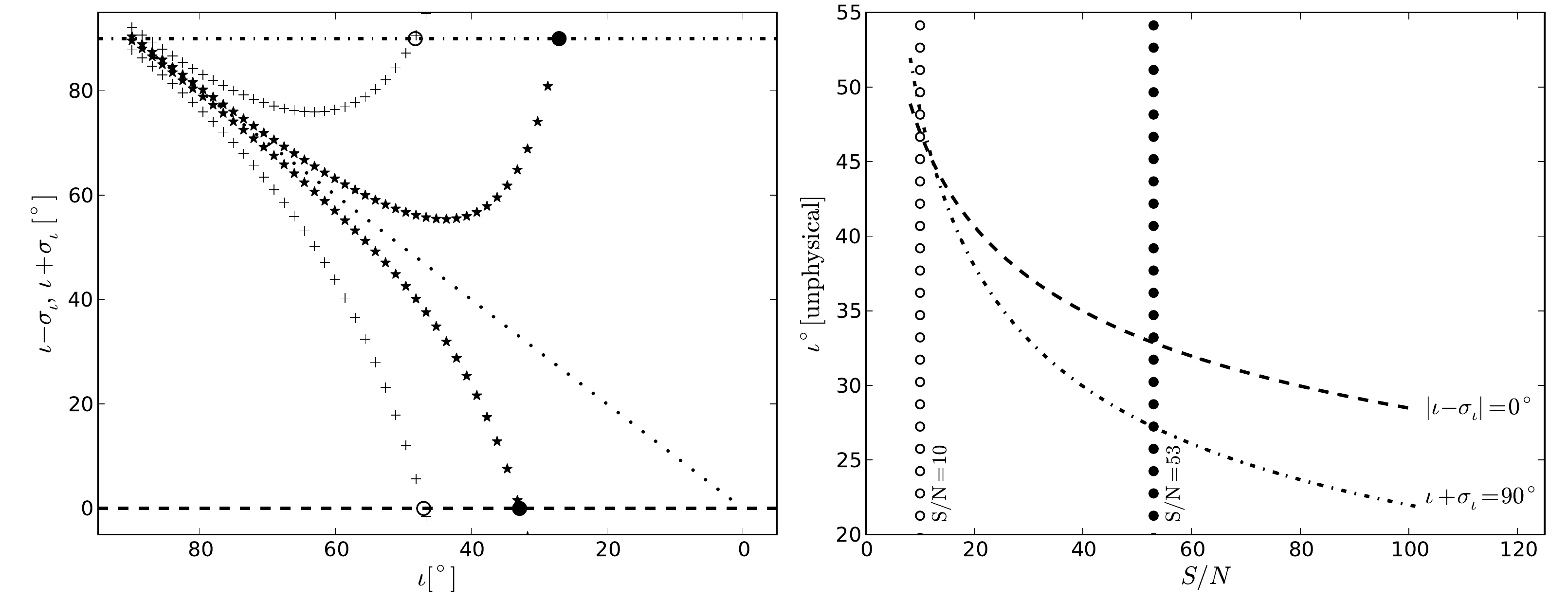}
\caption{\label{fig:unphys_di} Left: 1-$\sigma$ uncertainty in the inclinations
  vs. source inclinations for two values of S/N: plusses (outer curves)
  correspond to S/N=10, stars (inner curves) correspond
  to S/N=53. The curves above the dotted line correspond to $\iota
  +\sigma_{\iota}$, and those below to $\iota
  -\sigma_{\iota}$. The thick horizontal dashed and dashed-dotted lines are the
  physical limits of $\iota$: below the line of $\iota\pm
  \sigma_{\iota}=0^{\circ}$ and above 
  the line of $\iota\pm \sigma_{\iota}=90^{\circ}$, $\sigma_{\iota}$
  can be truncated, since that value of $\iota$ is bounded 
  between $[0^{\circ}, 90^{\circ}]$.  Right: `Unphysical $\iota$' versus S/N
  for the systems from MC1. The thick dashed-dotted line 
  calculated from the condition $\iota -\sigma_{\iota} = 0^{\circ}$, and the
  thick dashed line from the condition $\iota +\sigma_{\iota} =
  90^{\circ}$ demonstrated in the left panel. At an S/N of 10, the 
 in open circles show a lower limit and the filled
   circles show an upper limit at the
  intersections with the two curves from the conditions above.}
\end{figure*}
Consider the $\mathcal{A}-\iota$ error ellipse for a given signal $h$,
with true signal values $\iota_0$ and $\mathcal{A}_0$, where the two
parameters are strongly correlated and where $\sigma_{\iota} >> \pi$. 
The corresponding accuracy $\sigma_{\mathcal{A}}$ is then also highly
inflated.  Now consider a second signal $h(\mathcal{A}_1,\iota_1)$,
inside the error ellipse, but far from the true values, and such that
$\iota_1$ = $\iota_0 + 2n\pi$ with $n > 1$.  From its cyclic
behaviour, $h(\mathcal{A}_1,\iota_1) = h(\mathcal{A}_1,\iota_0)$
should hold. However, this last point would typically lie far outside
the error ellipse in the $\mathcal{A}-\iota$ plane, which would
indicate that this is actually a signal that is very \emph{dissimilar}
to the true signal. Hence, the fact that the inclination is physically
limited to the range $[0,\pi]$ and that this is not
taken into account in the FIM method means that we should not only
limit the uncertainty in inclination to its physically allowed range,
but that we should limit the uncertainty in amplitude
accordingly. Specifically, the corrected amplitude 
uncertainty, $\sigma_{\mathcal{A}}'$, is given by, 
\begin{equation}
\sigma_{\mathcal{A}}' = \sigma_{\mathcal{A}}[\mathrm{VCM}] \times
\frac{\sigma_{\iota}'}{\sigma_{\iota}[\mathrm{VCM}]},
\end{equation} 
where $\sigma_{\iota}'$ is the corrected inclination uncertainty. 

\begin{figure}[!h]
\centering
\includegraphics[width=\columnwidth]{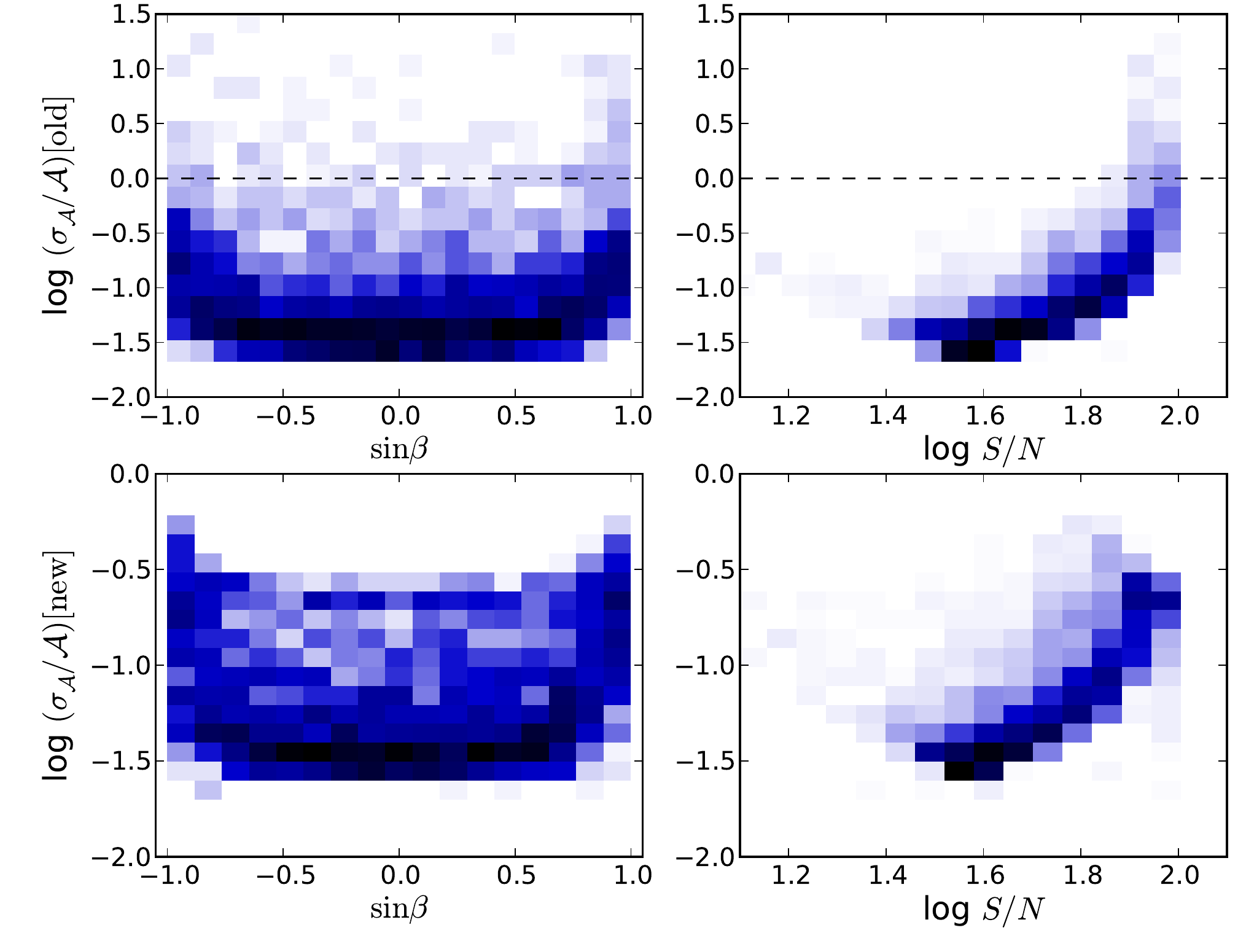}
\caption{\label{fig:corrected} Corrections applied to the
  uncertainties in amplitude and inclination from
  Figure~\ref{fig:errs_beta}. The row titled `new' was produced
  using the curves from the right panel of the
  Figure~\ref{fig:unphys_di} on MC1 simulations.}
\end{figure}

The value of inclination where its uncertainty becomes unphysical can
occur at two places, i.e. where $\iota +\sigma_{\iota} > \pi/2$
\footnote{Since $\iota$ is symmetric about $\pi/2$, we only considered
  the inclinations in the range $[0,\pi/2]$.}, and/or $\iota
-\sigma_{\iota} < 0$, as shown in the left panel of
Figure~\ref{fig:unphys_di}. In the panel, the 1-$\sigma$
uncertainty ranges of the inclination are plotted against the sources'
inclination for two cases of constant S/N of 10 and 53. The thick horizontal
dashed and dashed-dotted lines indicate the conditions where $\sigma_{\iota}$
becomes unphysical. At a fixed S/N of 10 (in plusses), these values
are at $\iota \sim 48^{\circ}$ and $\iota \sim 47^{\circ}$, 
respectively, as indicated by the open circles, whereas for a higher
S/N of 53 (in stars) the above conditions for $\sigma_{\iota}$
occur at $\iota \sim 27^{\circ}$ and $\iota \sim 32^{\circ}$ indicated
by filled circles. We calculated the inclinations at which these
intersections occur for a range of S/Ns, $10-110$ and show the
results in the right panel of Figure~\ref{fig:unphys_di}. These curves
can be used to apply corrections to the limits of $\sigma_{\iota}$ in
Figure~\ref{fig:errs_beta} from MC1, which in turn can be  
used to limit $\sigma_{\mathcal{A}}$ from Eq. 3. In
Figure~\ref{fig:corrected} we compare the uncertainties in 
$\mathcal{A}$ corrected in this way with those from
Figure~\ref{fig:errs_beta}. The corrected values for
$\sigma_{\mathcal{A}}$ differ for the high S/N systems, as can been
seen by comparing the two right-hand panels in
Figure~\ref{fig:corrected}. In particular, the tail at the large
$\sigma_{\mathcal{A}}/\mathcal{A}$ present in the top panel is absent
from the bottom panel.

There is a caveat to applying these corrections. In the left panel of
Figure~\ref{fig:unphys_di} we can understand the cut-off at
$0^{\circ}$ at the bottom of the figure. However, 
at the top, the cut-off of $90^{\circ}$ is not so clear. A system with
a lower inclination (e.g. $20^{\circ}$) cannot have an
error larger than $\sim 50^{\circ}$ because a system with an inclination
in the range $\sim 80^{\circ} - 90^{\circ}$ is distinguishable from a
system with an inclination of $\sim 0-60^{\circ}$, as has been pointed out
in Paper I. Thus the cut-off of $\iota+\sigma_{\iota} = \pi/2$ for lower
inclinations is a conservative estimate. The estimated posterior probability
distribution functions (PDFs) obtained using the FIM do not
predict the real PDFs for lower inclinations and lower S/Ns. Thus a
full Bayesian analysis that takes into account the limits of the
cyclic parameters as priors is needed to fully determine the PDF. This
is beyond the scope of this paper, and we used 
these (probably) over-estimated uncertainties to give the
corresponding bounds in the amplitude. 

In the same manner that $\sigma_{\mathcal{A}}$ was corrected in
Figure~\ref{fig:corrected}, we corrected for the MC2
calculations. The predicted improvement factors of the amplitude in the
bottom-right panel of Figure~\ref{fig:7X7_4X4} already include the
corrections mentioned above. 

\subsection{Use of FIM at \textit{S/N} $\sim 15$}
Another problem regarding the trustworthiness of the Fisher
matrix is the limit of the S/N at which one can trust the
values from VCM matrix. This has been addressed in 
\cite{1998PhRvD..57.4588N}, where the authors have performed Fisher
matrix studies in 
predicting the uncertainties of the parameters of the coalescing
binaries. They have compared these predictions with calculations from
Bayesian uncertainties, and their main finding is that above an S/N of 15, the
uncertainties from the Bayesian
calculations converge to Fisher matrix predictions. At an S/N of
10, the analytic predictions of Fisher matrix deviate by about
$6 \% $ and at a lower S/N of 7, they increase to about
$25 \%$. For MC1, the lowest S/N we considered is above $\sim 15$, 
as indicated in Figure~\ref{fig:SN_params}, and consequently the results
following MC1 are reasonable. However, real observations for eLISA
will of course also contain lower S/Ns. 

\section{Conclusion}
We presented a follow-up study of correlations between parameters that
characterise Galactic binaries that will be observed by eLISA in the
frequency range $10^{-4} \:\mathrm{Hz}<f<1 \:\mathrm{Hz}$. In Paper I
we investigated the correlations between parameters by considering 
three of the verification binaries (which are already known optically
and are thus guaranteed sources for eLISA), and we explained these
correlations and parameter uncertainties as a function of the
inclination of the system. In this paper, we addressed the effect
of all position and orientation parameters on the uncertainties
and the possible strong correlations between them. We found that the
overall effect is dominated by sky position and explained that
dependence. Our main conclusions are: 
\begin{enumerate}
\item After inclination, the ecliptic latitude of the source has the
  strongest influence on its parameter uncertainties, mainly due to
  its dependence on the signal-to-noise ratio and the correlations
  that can become very strong at the poles. For a given source
  distance, the parameter uncertainties (except for
  $\sigma_{\beta}$) are larger for a source at the
  ecliptic pole than for a source at the ecliptic plane.
  This is because a signal received from a source at the plane is
  stronger than that from a source at the pole.
\item In general, the uncertainty in sky location $\sigma_{\Omega}$ is
  large, i.e. in the order of square degrees. By constraining
  $\lambda$ and $\beta$
  from optical data, the uncertainties in $\psi$, $\phi$, $\iota$, and
  $\mathcal{A}$ in the GW analysis can be improved by a factor of
  $\sim 2$. This factor 
  is dependent on the inclination and the ecliptic latitude of the system.
\item By constraining $\iota$ in addition to the sky location, the
  improvement in $\sigma_{\mathcal{A}}$ is remarkably large (factors of
  10 depending on the S/N of the source). Systems with $\iota =
  30^{\circ}$ can have improvement factors as high as 60. 
\item The remaining angular parameters $\lambda$ and $\psi$ do not
  influence the correlations or uncertainties significantly.
\item The uncertainties in amplitude for low-inclination systems for
  which the corresponding uncertainties in the inclinations are
  unphysical (i.e. beyond their physical bounds) are also 
  exaggerated. Fisher matrix predictions for these uncertainties
  have to be corrected for such systems. 
\end{enumerate}

\begin{acknowledgements}
This work was supported by funding from FOM. We are very grateful to
Michele Vallisneri for providing support with the \textit{Synthetic LISA} and
\textit{Lisasolve} softwares. 
\end{acknowledgements}
\bibliographystyle{aa} 
\bibliography{/Users/swetashah/Documents/writings/literature_data_analysis,/Users/swetashah/Documents/writings/literature_binary_science}

\begin{appendix}
\section{Correlations as a function of $\beta$}
We found that several correlations between source parameters become
amplified at certain ecliptic 
latitudes and thus are a strong function of $\beta$, as shown in
Figure~\ref{fig:corrs_vs_elat}. Several of these correlations do not have
preference on the inclination of the system, and therefore we show a 2d
histogram to show the dependency on the ecliptic latitude.
In the figure we show (only) the correlations that become strong,  
i.e. where the normalised correlation $|c_{ij}| > 0.5$. Some of
these correlations have been explained in Paper I and are only
dependent on the inclination of the system; we show them here to
confirm that claim. The remaining correlations and their dependency on
the latitude can be understood by considering the orientation of
the binary relative to the detector. We comment on or explain each of
them below, 
\begin{enumerate}
\item $c_{\mathcal{A}\iota}$: Explained in Section 3.1. 

\item $c_{\phi\psi}$ and $c_{\phi f}$: Similarly to
  $c_{\mathcal{A}\iota}$,  these are strongly influenced by the
  source's inclination. The correlation $c_{\phi\psi}$ in the
  bottom-left panel has a distribution that looks
  very similar to the case of $c_{\mathcal{A}\iota}$ and here
  too, phase and polarisation are strongly correlated for face-on and 
  mildy face-on systems, since these two angles describe similar
  properties. For edge-on systems, the two angles are defined on 
  two planes that are perpendicular to each other, making them 
distinct. The correlation $c_{\phi f}$ in the top-right
  panel is strong for edge-on
  systems because a small shift in phase is very similar to a small
  change in frequency. However, for face-on systems, the phase
  and frequency become degenerate owing to the uncertainty in
  $\phi$, which is inflated due to the strong correlation with
  $\psi$. There is a slight dependency of this correlation on
  $\sin\beta$ for the edge-on orientations since the correlation has a
  larger spread at the ecliptic poles. This is due to the
  spread of the $c_{\phi\psi}$ for $\iota > 80^{\circ}$ systems where
  the spread in the phase at the poles also inflates the frequency. However, 
  $\iota < 60^{\circ}$, $c_{\phi\psi}$ for has no such spread, and thus
  the distribution of $c_{\phi\psi}$ vs. $\sin\beta$ for these systems
  lies in a narrow band throughout. 

\item $c_{\psi\lambda}$: This can be 
understood by considering a binary with $\iota =
90^{\circ}$ close to one of the poles; rotating the binary about its
line-of-sight is equivalent to changing its $\psi$ and $\lambda$ because they are
defined on the same plane (perpendicular to the
line-of-sight). This also holds for a face-on binary, and thus if a
source is close to the poles, $\psi$ and $\lambda$ can be highly
correlated. The spread of this correlation is most likely due to the
fact that $\lambda$ is not well defined at the poles and thus the
inflated uncertainty in the longitude implies a spread of correlations
at the poles.

\item $c_{\phi\lambda}$: This correlation is explained by the
  combination of $c_{\psi\lambda}$ and $c_{\phi\psi}$, which are
  explained above. Since $\psi$ and $\lambda$ can be
  highly correlated at the poles and $\phi$ and $\psi$ are strongly
  correlated for face-on systems, we found that $\phi$ and
  $\lambda$ should also be highly correlated at the poles, at least for
  the systems with (mildly) face-on orientations. For the edge-on systems,
  this strong correlation can also exist because even though
  $c_{\phi\psi}$ is not as strong in this case, the values of this correlation (as
  pointed out above) spread at the poles for $\iota=90^{\circ}$. This
  spread is responsible for the spread of $c_{\psi\lambda}$ around
  the poles for edge-on systems.

\item $c_{\iota\lambda}$ and $c_{\mathcal{A}\lambda}$: A strong
  correlation between inclination and the longitude is intuitively
  not as clear. However, this correlation combined with 
  $c_{\mathcal{A}\iota}$ for face-on systems would explain the 
  correlation between amplitude and longitude, $c_{\mathcal{A}\lambda}$.

\item $c_{\phi\iota}$: Like above, the correlation between phase and inclination
  follows from the combination of $c_{\phi\lambda}$ and
  $c_{\iota\lambda}$, both of which are stronger close to the poles. 

\end{enumerate}

\end{appendix}

\end{document}